\documentclass[reprint, superscriptaddress, amsmath, amssymb, prl, floatfix]{revtex4-2}
\usepackage{graphicx}
\usepackage{bm}
\usepackage{amsmath}
\usepackage{xcolor}
\usepackage[colorlinks,linkcolor = blue, citecolor = blue,urlcolor = blue]{hyperref}
\usepackage{subcaption} 
\usepackage{physics}
\usepackage{amsmath}
\usepackage{amssymb}
\usepackage[font = small]{caption}
\usepackage{dsfont}
\usepackage{overpic}
\usepackage{siunitx}
\makeatletter
\long\def\@makecaption#1#2{%
  \par
  \vskip\abovecaptionskip
  \begingroup
    \small\rmfamily
    \flushing
    \@make@capt@title{#1}{#2}\par
  \endgroup
  \vskip\belowcaptionskip}
\makeatother

\begin{document}

\title{Extreme sensitivity of nonlinear trajectories enhances optical spectral broadening}

\author{Jiachen Wang*}
\thanks{These two authors contributed equally to this work.}
\affiliation{Department of Physics, The Pennsylvania State University, University Park, Pennsylvania 16802, USA}
\author{Koorosh Sadri*}
\thanks{These two authors contributed equally to this work.}
\affiliation{Department of Physics, The Pennsylvania State University, University Park, Pennsylvania 16802, USA}
\author{Mikael C. Rechtsman}
\email{mcrworld@psu.edu}
\affiliation{Department of Physics, The Pennsylvania State University, University Park, Pennsylvania 16802, USA}

\date{\today}
\begin{abstract}
    The generation of a wide spectrum of light in highly nonlinear optical fibers has broad application in spectroscopy, microscopy and medical imaging~\cite{dudley2006supercontinuum,dudley2010supercontinuum}.  To generate such a `supercontinuum', an ultrashort pulse of light is injected into a highly nonlinear fiber. Then, in a process called self-phase modulation, the nonlinearity of the fiber causes the spectrum to start broadening as the pulse becomes chirped during propagation, seeding further cascaded nonlinear processes.  The dynamics associated with supercontinuum generation are captured mathematically as a pulse profile evolving in time and occupying a single spatial mode.  Here, we theoretically and experimentally demonstrate that a waveguide composed of multiple coupled cores - a `photonic molecule'\cite{yuan2023soliton,zhang2021squeezed} - rather than just a single core, gives rise to greater self-phase modulation for a given input power.  This is perhaps counterintuitive because it may be naively expected that the strongest nonlinear effects would be achieved by concentrating all optical power in one waveguide. The increased broadening arises due to the extreme sensitivity of trajectories near a separatrix of the nonlinear dynamics.  This sensitivity leads to a distortion of the temporal shape of the pulse, resulting in a broader spectrum.  The effect is reminiscent of the sensitivity associated with exceptional points in coupled-resonator systems, but does not suffer in the same way from the parasitic effects of noise.  This suggests that by including multiple cores, a straightforward modification of conventional nonlinear fiber design, supercontinuum sources seeded by self-phase modulation can generate a significantly wider spectrum.  More broadly, this demonstrates that the sensitivity to initial conditions associated with nonlinear dynamics may be utilized to generate stronger nonlinear effects in optics.    
\end{abstract}

\maketitle

The use of optical fibers to generate a broadband ``supercontinuum'' spectrum of light from a relatively narrowband pulse is essential in an array of scientific and technological settings, including broadband spectroscopy, microscopy, and optical coherence tomography for medical imaging.  Supercontinuum generation is highly complex, involving multiple nonlinear cascaded processes, but it starts simply: when a pulse is injected into a guided optical mode, the Kerr nonlinearity of the glass causes a distortion of the phase profile of the pulse, causing the effective frequency at early times to be lower than that of the later times.  This `chirping' of the pulse, called self-phase modulation, acts to launch the complex nonlinear dynamics of supercontinuum generation.  It has been nearly exclusively studied as a single-spatial mode process, with the focus of research being on the complex nonlinear dynamics in time.

In the domain of optical sensing with high-Q photonic resonators \cite{vollmer2008whispering}, it has been proposed~\cite{wiersig2014enhancing} that by using multiple coupled resonators instead of just one, extreme sensitivity to external perturbations could be realized near exceptional points of the associated driven-dissipative open-system Hamiltonians~\cite{chen2017exceptional,hodaei2017enhanced}.  However, this also leads to the enhancement of noise in the output signal associated with the steady state, mitigating the increased sensitivity of the exceptional point per se~\cite{lau2018fundamental}.  In coupled waveguide systems~\cite{christodoulides1988discrete, eisenberg1998discrete, eisenberg2000diffraction,christodoulides2003discretizing,fleischer2003observation} the dynamics are different in the sense that they are conservative, not driven-dissipative: once a pulse is injected at the input facet, it evolves with the distance of propagation along the waveguide axis, and never reaches a steady state.  Provided that the optical power is sufficiently high for significant nonlinearity to manifest, the resulting dynamics can be very rich, with effects such as nonlinear bifurcations~\cite{Eilbeck1985, kevrekidis2009discrete}, spatial soliton formation~\cite{eisenberg1998discrete, fleischer2003observation}, and modulation instability~\cite{hasegawa1984generation,kivshar1992modulational}.  Under certain conditions, nonlinear trajectories can be extremely sensitive to initial conditions, even if they are not chaotic, as a result of separatrices that define a boundary between dramatically different orbits.  

Here, we propose and demonstrate that the extreme sensitivity, such as that associated with separatrices in nonlinear dynamical trajectories, can lead to a significant enhancement in self-phase modulation compared with a single waveguide. The essence of the broadening mechanism is that in a system composed of more than one waveguide, the nonlinear dynamics can exhibit sharp separatrices, and if the peak of the pulse lies on one side of the separatrix and the tails on the other, the pulse will be significantly distorted, resulting in spectral broadening.  We demonstrate this effect using devices consisting of one, two and three coupled waveguides, which are fabricated with the femtosecond direct laser-writing method~\cite{doi:10.1126/science.aba8725,jurgensen2021quantized}.  We characterize the spatiotemporal dynamics of light propagation by measuring the spatially-resolved spectrum at each waveguide output.  We then compare our results with numerical simulations of the predicted dynamics.  The multiple coupled waveguides constitute a `photonic molecule', which have been used previously in realizations based on coupled resonators to enhance nonlinear effects, despite the fact that optical power must be spread out among multiple elements~\cite{yuan2023soliton,zhang2021squeezed}. 

\begin{figure} [h!t]
    \centering
    \begin{overpic}[width = 1 \linewidth]{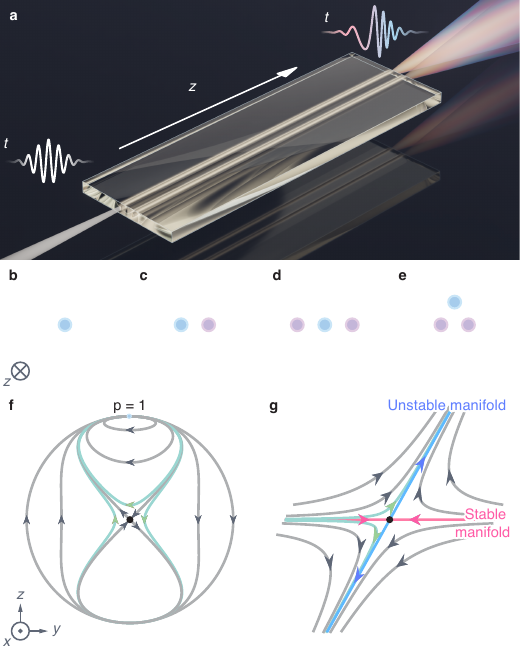} 
    \end{overpic}
    \caption{ \textbf{a}, Schematic illustration of the coupled waveguide system, where a pulse of light is injected into one waveguide and different temporal components of the pulse tunnel to another one through evanescent coupling.  
    \textbf{b--e}, Schematic illustration of the four waveguide geometries studied here, with the excited site highlighted in blue: \textbf{b}, single isolated waveguide; \textbf{c}, a horizontal pair; \textbf{d}, a horizontal open-boundary trimer; and \textbf{e}, a closed-boundary triangular trimer. \textbf{f}, The dynamical paths on the Bloch sphere for a symmetric two waveguide system with single-site initial condition, i.e., at the north pole ($p = +1$, marked by a blue dot). Low power paths $r < 4J/g$ reach all the way to the south pole at $p = -1$, while high power paths ($r > 4J/g$) stay above the equator at $z = 0$. The separatrix, corresponding to $r = 4J/g$, asymptotically approaches a hyperbolic fixed point at $y = z = 0$ (marked by a black dot). \textbf{g}, The linearized dynamics close to a generic hyperbolic fixed point. The existence of a stable direction (highlighted with pink), allows for paths close to the separatrix to linger around the fixed point for a longer time while being pushed exponentially farther apart along the unstable direction (highlighted with blue).}
    \label{fig:1}
\end{figure}

Our system consists of a set of single-mode optical waveguides fabricated in Corning Eagle XG glass. They are positioned sufficiently close to one another to enable evanescent coupling between neighboring waveguides (see Fig.~\ref{fig:1}a). Within the paraxial approximation, the slowly varying field amplitude $\psi_i(t,z)$ in waveguide $i$ obeys the discrete nonlinear Schr\"odinger equation. In our experiments, the input spectrum is sufficiently small that the dispersion length greatly exceeds the sample length, $L_D \gg L_{\mathrm{device}}$, allowing temporal dispersion to be neglected. Since dispersion acts to effectively couple different time slices in the pulse, the dynamics therefore reduces to the temporally decoupled discrete nonlinear Schr\"odinger equation:
\begin{equation}\label{GPE}
    i\pdv{}{z}\psi_i(t,z) = -\sum_j J_{ij}\psi_j - g|\psi_i|^2\psi_i,
\end{equation}
where $z$ denotes the propagation distance, $t$ denotes time, $J_{ij}$ are the evanescent coupling constants between waveguides $i$ and $j$, and $g$ is the Kerr nonlinearity coefficient.

The input pulse determines the initial condition $\boldsymbol{\psi}(t,0) = \boldsymbol{\psi}^{\mathrm{in}}(t)$. In our experiments, the input beam excites a single waveguide, so $\psi_i^{\mathrm{in}}(t) = \delta_{i,i_0}f(t)$, where $i_0$ is the injection site and $f(t)$ is the temporal pulse shape. We let $\Phi$ denote the dynamical map generated by Eq.~\eqref{GPE} from $z = 0$ to $z = Z$; such that the output pulse is
\begin{equation}
    \boldsymbol{\psi}^{\mathrm{out}}(t) \equiv \boldsymbol{\psi}(t,Z) = \Phi(\boldsymbol{\psi}(t, 0)) = \Phi\!\left(\boldsymbol{\psi}^{\mathrm{in}}(t)\right).
\end{equation}

Although the input $\boldsymbol{\psi}^{\mathrm{in}}(t)$ is a smooth temporal pulse, the nonlinear dynamical map $\Phi$ can exhibit strong sensitivity to initial conditions. In particular, small variations in the input temporal envelope may be strongly amplified by a large derivative of the dynamical map, producing sharp temporal features in the output pulse. Since sharper temporal features correspond to enhanced high-frequency Fourier components, this mechanism substantially increases the spectral weight in the tail of the output spectrum.

To quantify this sensitivity, we introduce the linearized evolution operator $T$, defined by the first-order variation of the output with respect to the input, or equivalently, the derivative of the dynamical map $\Phi$; such that $(\Re \delta \boldsymbol{\psi}^{\mathrm{out}}, \Im \delta \boldsymbol{\psi}^{\mathrm{out}}) = T(\Re \delta \boldsymbol{\psi}^{\mathrm{in}}, \Im \delta \boldsymbol{\psi}^{\mathrm{in}})$. The operator norm $\|T\|$, coinciding with the largest singular value of $T$, characterizes the maximal amplification of infinitesimal perturbations between input and output configurations. In the absence of nonlinearity, where the dynamical map $\Phi$ is a unitary transformation, the norm $\|T(z)\| = 1$ holds for all propagation distances signifying no amplified sensitivity to initial conditions. In regimes where the dynamics exhibits exponential sensitivity to initial conditions, the growth of $\|T(Z)\|$ is governed by a Lyapunov exponent $\lambda$ such that $\|T(Z)\|\sim e^{\lambda Z}$; and thus, $\lambda > 0$ corresponds to exponential amplification of output variations and consequently strong enhancement of the spectral tail.

Here we study four geometries, all with single-site excitation as shown in Fig.~\ref{fig:1}b--e: a single isolated waveguide (Fig.~\ref{fig:1}b); a horizontal pair excited on one site (Fig.~\ref{fig:1}c); a horizontal three-waveguide open-boundary trimer excited at the central site (Fig.~\ref{fig:1}d); and a triangular trimer consisting of an equilateral triangle, excited at the apex site (Fig.~\ref{fig:1}e). We show that all three multi-waveguide geometries can exhibit excess spectral broadening compared to the standard self-phase modulation present in a single waveguide system and that this phenomenon is accompanied by enlarged dynamical sensitivity as measured by the operator norm $\|T\|$.

\begin{figure*}[h!t]
    \centering
    \begin{overpic}[width = 1 \linewidth, grid = F]{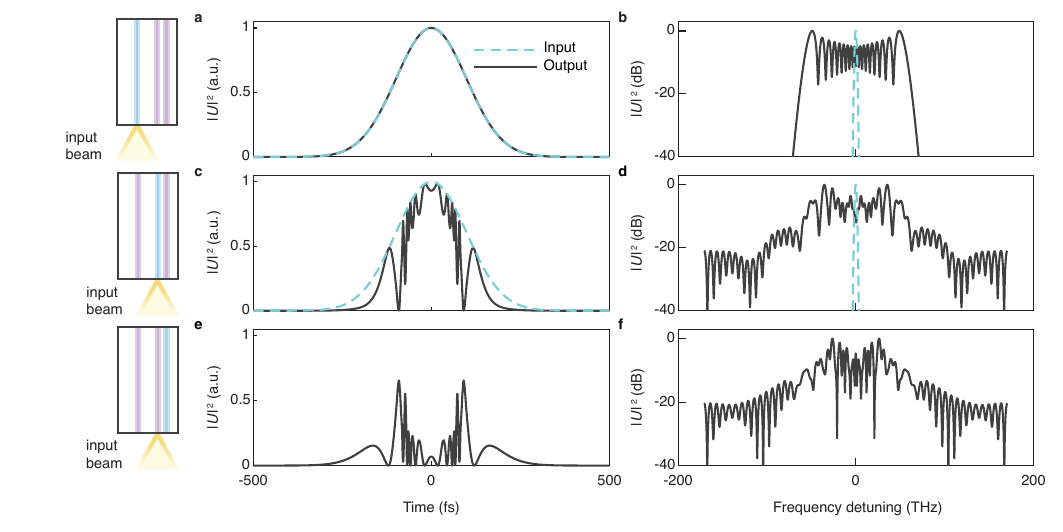} 

    \end{overpic}
    \caption{Simulated pulse profiles in the time domain (\textbf{a},\textbf{c},\textbf{e}) and frequency domain (\textbf{b},\textbf{d},\textbf{f}). Dashed lines show the input pulse, and solid lines show the output after propagating through the corresponding waveguide. \textbf{a},\textbf{b}, Single waveguide. \textbf{c},\textbf{d}, Waveguide of the pair in which the pulse was injected. \textbf{e},\textbf{f}, Neighboring waveguide of the pair, into which there was no input. Schematic illustrations on the left indicate each geometry; yellow triangles mark the input waveguide site, and the blue highlight indicates the waveguide whose temporal and spectral profile is plotted.}
    \label{fig:2}
\end{figure*}

For a single waveguide, the dynamical map for Eq. \eqref{GPE} is found analytically as $\Phi_{\mathrm{single}}(\psi) = e^{igZ|\psi|^2}\psi$. The sensitivity is also straightforward to compute as $\|T(Z)\| = gZ|\psi|^2 + \sqrt{1 + g^2Z^2|\psi|^4}$, which exhibits only linear growth at large propagation distances. Consequently, a single waveguide cannot generate exponential sensitivity. For a Gaussian input pulse, the output merely acquires a nonlinear chirp while maintaining the same amplitude profile of the pulse, resulting in spectral broadening (See Fig.~\ref{fig:2}a,b). This is standard self-phase modulation (SPM)~\cite{boyd2008nonlinear}.

We next consider the case of two coupled waveguides, an integrable Hamiltonian system whose phase space structure is closely analogous to a simple pendulum, as well as oscillations of a Josephson junction \cite{josephson62, smerzi97, raghavan99}. Just as a pendulum's separatrix divides bounded oscillations (libration) from continual rotations around the pivot, the waveguide system features a critical separatrix trajectory that divides two qualitatively distinct regimes: low-power complete oscillations between waveguides and high-power partial self-trapping. For generic trajectories in both systems, the sensitivity operator grows only linearly with time or propagation distance. The unique exception occurs near the separatrix, which approaches a hyperbolic fixed point (the pendulum’s upright equilibrium) asymptotically. Trajectories initialized close to this boundary spend arbitrarily long times near this unstable equilibrium, where infinitesimal perturbations are exponentially stretched along its unstable manifold, generating an anomalously large sensitivity operator. Consequently, as temporal slices of an input pulse cross this critical threshold, nearby time slices become highly sensitive to small amplitude variations. This localized sensitivity generates sharp temporal features in the output pulse and significantly enhances the high-frequency spectral weight, resulting in a substantially heavier tail in the output spectrum when compared to a single waveguide (see Fig.~\ref{fig:2}c--f).

To analyze the dynamics of the two-waveguide system more precisely in the language of a pendulum, we parametrize the wavefunction as $\psi_1 = e^{i(\alpha - q/2)}\sqrt{r(1 + p)/2}$, and $\psi_2 = e^{i(\alpha + q/2)}\sqrt{r(1 - p)/2}$ where $r = |\psi_1|^2 + |\psi_2|^2$ is proportional to the conserved total power and the overall global phase, $\alpha$ is non-dynamical. If $J_{12} = J_{21} = J$, and $-J_{11} = J_{22} = V$, then the conjugate variables $p$ and $q$ follow the dynamics induced by the Hamiltonian $H(q, p) = -2J\sqrt{1 - p^2}\cos(q) - 2V(1 - p) + rg(1 - p^2)/2$. For the symmetric case, $V = 0$, the dynamical path starting from the single site excitation point, $p = +1$, lies on the curve described by $H(q, p) = 0$. This curve is connected to the single site excitation on the other waveguide, $p = -1$, for $r < r_c = 4J/g$ and is constrained to the $p > 0$ region for $r > 4J/g$. The separatrix corresponding to $r = 4J/g$ connects to a hyperbolic fixed point at $q = p = 0$. Around this fixed point, the Hamiltonian takes the hyperbolic profile $H(q, p) \approx J(q^2 - p^2)$. In Fig.~\ref{fig:1}f, we have projected the dynamical trajectories starting from the single site excitation $(p = 1)$ corresponding to different $r$ values on the surface of the Bloch sphere. Fig.~\ref{fig:1}g, shows a close up profile of the hyperbolic fixed point that the separatrix ends in. A similar transition occurs for the asymmetric case where $V\neq 0$, although the critical nonlinearity is higher: $r_c(J, V) > 4J/g$.

For arrays containing more than two waveguides, the nonlinear dynamics are generically nonintegrable and may exhibit chaotic regions with positive Lyapunov exponent over a finite volume of phase space. In this regime, exponential sensitivity to initial conditions is no longer restricted to trajectories near a separatrix but occurs for entire families of trajectories. Consequently, once the input power exceeds the threshold for the onset of chaos, the sensitivity operator grows exponentially with propagation distance, thereby producing a strong enhancement of the spectral tail of the output pulse. 

To characterize the output spectrum in the experiment, we use a pulsed laser with tunable pulse duration (Amplitude Satsuma; wavelength $\SI{1030}{nm}$, repetition rate $\SI{5}{kHz}$), focus the beam onto a single input site, and collect the output from the same site for analysis with an optical spectrum analyzer (Anritsu MS9740A). The pulse duration before the sample is measured with an autocorrelator (A.P.E.\ PulseCheck). The Kerr nonlinear coefficient for the fabricated waveguides was measured to be $\gamma = \SI{7.3e-4}{(W.m)^{-1}}$, and the propagation length of the waveguides is $\SI{76.2}{mm}$. The adjacent waveguides are evanescently coupled, with horizontal hopping amplitude $J_{\rm{hor}} = 11.61\,\mathrm{mm}^{-1}\,e^{-\SI{0.276}{\per\micro\meter}\, a}$ and diagonal hopping amplitude between one base site and the apex site in the triangular trimer $J_{\rm{diag}} = 10.76\,\mathrm{mm}^{-1}\,e^{-\SI{0.256}{\per\micro\meter}\, a}$, where $a$ is the separation of neighboring waveguides. In the direct laser writing process, there may be significant detuning of the propagation constants in neighboring waveguides within a given device, due to the writing beam overlapping other waveguides. We experimentally measure this detuning using linear propagation measurements, and take these into account in numerical simulations.

\begin{figure*}[h!t]
    \centering
    \begin{overpic}[width = 1 \linewidth]{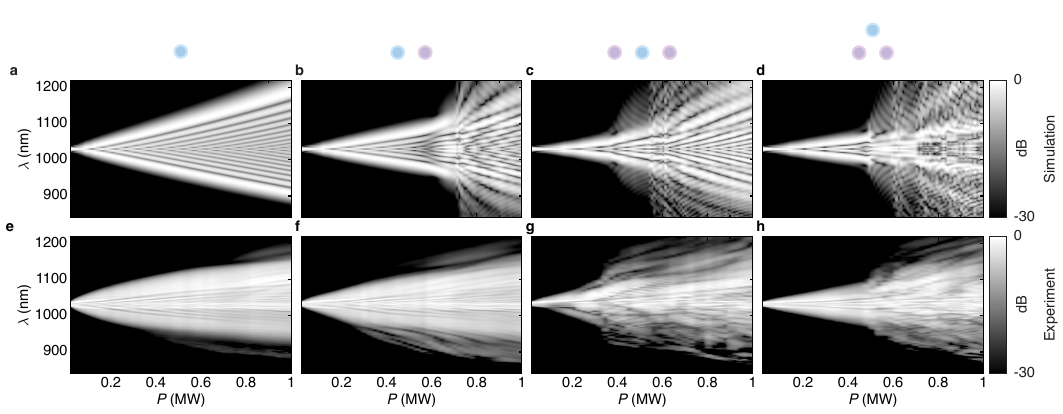} 
    \end{overpic}
    \caption{Simulated and experimental output optical spectra as a function of input peak power. \textbf{a--d}, Simulated output of a single isolated waveguide, a horizontal pair, a horizontal open-boundary trimer, and a closed-boundary triangular trimer, respectively. \textbf{e--h}, Experimental output corresponding to \textbf{a--d}. Schematic illustrations above each column show the corresponding geometry; the input waveguide, whose output is spectrum is shown here, is highlighted in blue. Waveguide separation $a = \SI{17}{\micro\meter}$ for all coupled geometries.}
    \label{fig:3}
\end{figure*}

Figure~\ref{fig:3} shows simulated and experimental output spectra as a function of peak input power $P$ for the four geometries with nearest neighbor separation $a = \SI{17}{\micro\meter}$ for the coupled geometries, and the input pulse full width at half maximum duration $\tau = 335$~fs. In simulation, the spectrum of a single waveguide broadens moderately as a result of SPM as seen in Fig.~\ref{fig:3}a. However, the spectrum of a coupled geometry first tracks the single-waveguide case, but broadens abruptly above a threshold. The threshold is near $\SI{0.8}{MW}$ for a horizontal pair excited on one site (Fig.~\ref{fig:3}b), near $\SI{0.4}{MW}$ for a horizontal three-waveguide open-boundary trimer excited at the central site (Fig.~\ref{fig:3}c), and near $\SI{0.5}{MW}$ for a triangular trimer consisting of an equilateral triangle, excited at the apex site (Fig.~\ref{fig:3}d). This abrupt change coincides with the onset of extreme sensitivity. The experimental data shown in Fig.~\ref{fig:3}e--h faithfully reproduce both the low-power SPM regime and the threshold-like onset of broadening seen in simulation, validating the model parameters. 

\begin{figure*}[h!t]
    \centering
    \begin{overpic}[width = 1 \linewidth, grid = F]{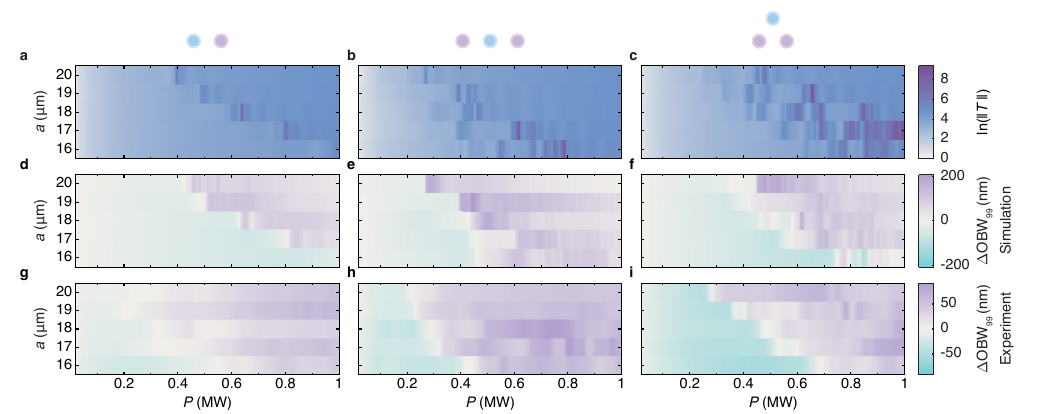} 
    \end{overpic}
    \caption{Excess output spectral broadening relative to a single waveguide due to enhanced sensitivity to initial conditions, as a function of peak power $P$ and waveguide separation $a$. \textbf{a--c}, The sensitivity measure characterized by the largest singular value of the derivative operator of the dynamical map, $\|T\|$, of a horizontal pair, a horizontal open-boundary trimer, and a closed-boundary triangular trimer, respectively. \textbf{d--f}, Simulated values for the excess spectral broadening measured by the differential 99\% bandwidth $\Delta\rm{OBW}_{99}$ of the corresponding structures. \textbf{g--i}, Experimential results corresponding to \textbf{d--f}. Schematic illustrations above each column show the corresponding geometry; the input waveguide, whose output is measured, is highlighted in blue.}
    \label{fig:4}
\end{figure*}

We now demonstrate the close relationship between the sensitivity signature $\|T\|$ and spectral broadening. To quantify spectral width, we use the 99\% occupied bandwidth ($\mathrm{OBW}_{99}$)—defined as the wavelength range containing 99\% of the spectral power, with 0.5\% excluded on either side~\cite{ITUR_SM328, ITUR_SM443, FCC_2202}. Figure~\ref{fig:4} plots $\|T\|$ (Figs.~\ref{fig:4}a--c) alongside simulated (Figs.~\ref{fig:4}d--f) and experimental (Figs.~\ref{fig:4}g--i) values for the excess output bandwidth, $\Delta\mathrm{OBW}_{99} = \mathrm{OBW}_{99}^\mathrm{coupled} - \mathrm{OBW}_{99}^\mathrm{single}$, across peak power $P$ and waveguide separation $a$ for a fixed input pulse duration ($\tau = 335$~fs). At lower powers, $\Delta\mathrm{OBW}_{99}$ remains near zero or weakly negative because the coupled system lacks enhanced nonlinear sensitivity. However, across all geometries and separation distances, an large peak in $\|T\|$ directly coincides with the onset of enhanced spectral broadening. This strong agreement between the experimental, simulated, and theoretical trends confirms that nonlinear sensitivity drives the enhanced spectral broadening observed here.

\begin{figure*}[h!t]
    \centering
    \begin{overpic}[width = 1 \linewidth, grid = F]{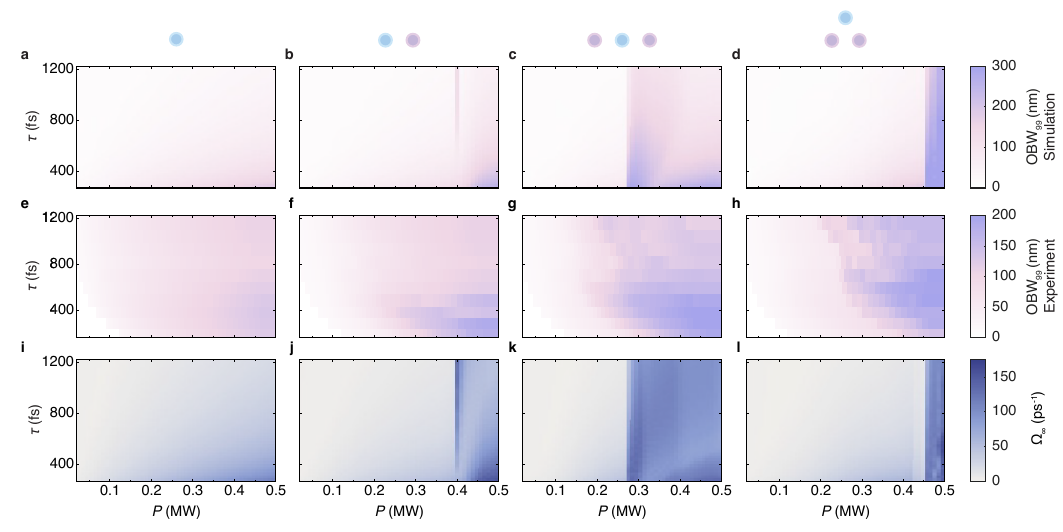} 

    \end{overpic}
    \caption{Output spectral broadening and the normalized maximum time derivative of the output pulse as a function of peak power $P$ and pulse duration $\tau$ for different waveguide geometries. \textbf{a--d}, Simulated values for the occupied bandwidth, $\mathrm{OBW}_{99}$ of a single isolated waveguide, a horizontal pair, a horizontal open-boundary trimer, and a closed-boundary triangular trimer, respectively. \textbf{e--h}, Experimental data for the occupied bandwidth, $\mathrm{OBW}_{99}$, corresponding to subfigures a--d. \textbf{i--l}, The normalized maximum output time derivative $\Omega_\infty$ (cf. Eq.~\eqref{OmegaInfty}) corresponding to the geometries of subfigures a--d. Schematic illustrations above each column show the corresponding geometry; the input waveguide, whose output is measured, is highlighted in blue. Waveguide separation $a = \SI{20}{\micro\meter}$ for all panels.}
    \label{fig:5}
\end{figure*}

In addition to the peak power, we may vary the duration of the input pulse without changing its spectrum by introducing a chirp in its phase profile; the chirp and stretching of the pulse is carried out dispersively using gratings. The top row in Fig.~\ref{fig:5}, consisting of panels a--d, shows simulation results of the dependence of the bandwidth, $\rm{OBW}_{99}$, on both the peak power as well as the pulse duration for the four waveguide geometries. The middle row, consisting of panels ~\ref{fig:5}e--h, shows the corresponding experimental results. In contrast to Fig.~\ref{fig:4}, here we plot the absolute bandwidth $\mathrm{OBW}_{99}$ for each geometry individually rather than the differential $\Delta\mathrm{OBW}_{99}$.  We clearly observe a jump in spectral broadening as a function of peak power for all pulse durations: this is directly associated with the separatrix in the dynamical map. The jump is sharper in numerical calculations as compared with experiment; this is likely due to the effects of loss and higher-order nonlinear effects that are neglected here. Furthermore, the spectral broadening also decreases with pulse duration in all cases, particularly those with multiple waveguides involved.  We explain the mechanism underlying this below, in terms of sensitivity to initial conditions and the separatrix associated with the dynamical map.  

Neighboring time slices within the pulse experience different amplification in their initial differences governed by the sensitivity operator $T$, which is a function of the amplitude of the pulse.  

Due to the U(1) phase symmetry present in Eq.~\eqref{GPE}, the overall phases are preserved by the dynamical map, indicating no enhanced sensitivity due to the phase structure of the chirp. In other words, two adjacent dynamical paths distinguished only by a difference in their overall phase exactly maintain their phase difference while otherwise following the same trajectory. In contrast, dynamical paths that start with different initial amplitudes can have their separation amplified as they evolve according to Eq.~\eqref{GPE}.  It is precisely this divergence at neighboring time slices that lead to rapid changes in the output pulse, in turn giving rise to pulse distortion and spectral broadening.  Therefore, since shorter pulses have steeper changes in amplitude as a function of time, we see greater spectral broadening at shorter pulse durations, explaining the vertical trend in each of Fig.~\ref{fig:5}a--h.  

Mathematically speaking, the input and output temporal derivatives satisfy $d\boldsymbol{\psi}^{\mathrm{out}}(t)/dt = T(\psi^{\mathrm{in}}(t))\, d\boldsymbol{\psi}^{\mathrm{in}}(t)/dt$. It follows that $|d\boldsymbol{\psi}^{\mathrm{out}}(t)/dt|\leq\|T(\psi^{\mathrm{in}}(t))\|\, |d\boldsymbol{\psi}^{\mathrm{in}}(t)/dt|$. Thus, the operator norm $\|T(\psi^{\mathrm{in}}(t))\|$ controls the maximum rate at which the output pulse may vary in time. Large values of $\|T(\psi^{\mathrm{in}}(t))\|$ therefore allow the nonlinear dynamics to generate extremely steep temporal variations from an otherwise smooth input pulse. The inequality is saturated if and only if the input temporal derivative is aligned with the right singular vector corresponding to the largest singular value of $T$. As explained earlier, changes in the pulse duration will lead to changes in this alignment and therefore end up varying the ultimate enhancement in the temporal derivative of the output pulse. In the bottom row of Fig.~\ref{fig:5}, panels i--l, we have plotted the magnitude of the steepest temporal derivative of the output pulse, normalized by the largest output amplitude for different peak powers and pulse durations. This is 
\begin{equation}\label{OmegaInfty}
    \Omega_\infty \equiv \frac{\max_t |d\boldsymbol{\psi}^\mathrm{out}(t)/dt|}{\max_t |\boldsymbol{\psi}^\mathrm{out}(t)|} = \frac{\|\dot{\boldsymbol{\psi}}^\mathrm{out}(t)\|_\infty}{\|\boldsymbol{\psi}^\mathrm{out}(t)\|_\infty}.
\end{equation}
This quantity, which measures the sensitivity associated with the dynamical map, strongly predicts the bandwidth for all four waveguide geometries as evidenced by the results in Fig.~\ref{fig:5}.  Clearly, the dynamical map separatrix causes a sharp increase in sensitivity that is associated with a dramatic increase in spectral broadening. 

In summary, we have demonstrated theoretically and experimentally an enhancement of self-phase modulation, the seed of supercontinuum generation, by using multiple coupled waveguides instead of just one -- namely, a `photonic molecule'.  The coupled-waveguide devices were fabricated using the femtosecond direct laser writing process, and experiments were carried out by injecting pulses into one of them. The mechanism proceeds via the exponential sensitivity associated with particular nonlinear dynamical trajectories that only appear when multiple spatial degrees of freedom are present.  In the simplest case of two equivalent waveguides, this sensitivity arises due to an unstable hyperbolic manifold that acts as a separatrix between two qualitatively different behaviors: complete oscillation between sites and partial self-trapping on one.  For the case of three coupled waveguides, increased sensitivity was also present but the mechanism thereof is more complex.  In all cases, we observed increased spectral broadening relative to the single-waveguide case that we showed to be associated with with the extreme sensitivity of the nonlinear trajectories.  While only the two and three waveguide cases were studied here, we expect that systems with more degrees of freedom can exhibit even greater spectral broadening, particularly when the dynamics enter the chaotic regime.  These results suggest that by including multiple cores in supercontinuum-generating fibers, a broader spectrum can be achieved at lower power.  More generally, we have shown that the extreme sensitivity associated with nonlinear dynamical trajectories can be used to significantly enhance nonlinear optical effects in systems with multiple degrees of freedom. \\

\noindent \textit{Acknowledgements}. The authors acknowledge the support of the Air Force Office of Scientific Research under the MURI program, agreement number FA9550-22-1-0339, as well as the Office of Naval Research under agreement number N00014-23-1-2102. The authors acknowledge Nicholas Smith of Corning Inc. for providing Eagle XG glass samples.

\bibliographystyle{naturemag}
\bibliography{Refs}
\end{document}